\documentclass{article}

\usepackage{arxiv}

\usepackage[toc,page]{appendix}

\usepackage{graphicx}
\usepackage{dcolumn}
\usepackage{bm}
\usepackage{appendix} 

\usepackage[mathlines]{lineno}

\usepackage[utf8]{inputenc}
\usepackage{gensymb} 
\usepackage{booktabs} 
\usepackage{subcaption} 
\usepackage{color} 
\usepackage{adjustbox}
\usepackage{amsmath}
\usepackage{mathtools, nccmath}
\usepackage{xparse}
\DeclarePairedDelimiterX{\set}[1]{\{}{\}}{\setargs{#1}}
\usepackage{listings}
\usepackage{enumitem}
\definecolor{codegreen}{rgb}{0,0.6,0}
\definecolor{codegray}{rgb}{0.5,0.5,0.5}
\definecolor{codepurple}{rgb}{0.58,0,0.82}
\definecolor{backcolour}{rgb}{0.95,0.95,0.92}
\usepackage{textcomp}

\usepackage{algpseudocode}
\usepackage{amsfonts}
\usepackage{lscape}
\usepackage{longtable}

\lstdefinestyle{mystyle}{
  backgroundcolor=\color{backcolour}, commentstyle=\color{codegreen},
  keywordstyle=\color{magenta},
  numberstyle=\tiny\color{codegray},
  stringstyle=\color{codepurple},
  basicstyle=\ttfamily\footnotesize,
  breakatwhitespace=false,         
  breaklines=true,                 
  captionpos=b,                    
  keepspaces=true,                 
  numbers=left,                    
  numbersep=5pt,                  
  showspaces=false,                
  showstringspaces=false,
  showtabs=false,                  
  tabsize=2
}

\usepackage[utf8]{inputenc}
\usepackage[english]{babel}
\usepackage{graphicx}
\usepackage{gensymb} 
\usepackage{booktabs} 
\usepackage{lipsum}

\usepackage[normalem]{ulem} 
\usepackage{bm}
\usepackage{bbm}
\usepackage{amssymb}
\usepackage{multirow}

\usepackage{url}
\usepackage[hidelinks]{hyperref}

\usepackage{algcompatible}
\usepackage{newfloat}
\DeclareFloatingEnvironment[
    fileext=loa,
    listname=List of Algorithms,
    name=ALGORITHM,
    placement=tbhp,
]{algorithm}
\newcommand{\beginsupplement}{%
        \setcounter{table}{0}
        \renewcommand{\thetable}{S\arabic{table}}%
        \setcounter{figure}{0}
        \renewcommand{\thefigure}{S\arabic{figure}}%
     }

\title{FLASH-enabled Proton SBRT for a challenging case of spine metastasis}

\usepackage{authblk}

\newcommand{\mycomment}[1]{}

\author[1]{Sophie Wuyckens}
\author[1]{Macarena Chocan Vera}
\author[2]{Rasmus Nilsson}
\author[2]{Viktor Wase}
\author[3]{Dario Di Perri}
\author[3]{Xavier Geets}
\author[1,4,5]{Edmond Sterpin}
\author[1]{John A. Lee}

\affil[1]{Université catholique de Louvain, Institut de Recherche Expérimentale et Clinique (IREC), Molecular Imaging, Radiotherapy and Oncology, Woluwe-Saint-Lambert, Belgium}
\affil[2]{RaySearch Laboratories AB, Stockholm, Sweden}
\affil[3]{Radiation oncology department, Cliniques Universitaires Saint-Luc, Brussels, Belgium}
\affil[4]{KULeuven, Department of Oncology, Laboratory of experimental radiotherapy, Leuven, Belgium}
\affil[5]{Particle Therapy Interuniversity Center Leuven - PARTICLE, Leuven, Belgium}

\renewcommand{\headeright}{Preprint}
\renewcommand{\undertitle}{Preprint}
\renewcommand{\shorttitle}{FLASH SBRT for a spine metastasis}

\begin{document}
\maketitle

\begin{abstract}
\textit{Objective.} The FLASH effect, characterized by potential sparing of organs at risk (OAR) through ultra-high dose rate irradiation, has garnered significant attention for its capability to address indications previously untreatable at conventional dose rates (DR) with hypofractionated schemes. While considerable biological research is needed to understand the FLASH effect and determine the FLASH modifying factors (FMF) for individual OARs, exploratory treatment planning studies have also emerged. This study aims to show that spinal metastases are candidate treatment sites likely to benefit from this phenomenon and establish the requisite FMFs to achieve the protective FLASH effect.  \\
\textit{Approach.}  A conformal FLASH Proton SBRT plan was generated for a patient with spine metastasis in a research version of RayStation11B (RaySearch laboratories AB, Stockhom) on an IBA Proteus Plus system. Two oblique posterior beams were used in the plan. The prescribed dose to the CTV was set according to 3 different fractionation regimens: 5 fractions (fx) of 7 Gy, 8 fx of 5 Gy, and 10 fx of 4.2 Gy. Spot filtering and sorting techniques were applied to maximize the 5\% pencil beam scanning DR in the spinal cord (SC). The FLASH effect was assumed to be observed within irradiated regions above 40 Gy/s and 4 Gy per fraction.\\
\textit{Result.} The generated plans successfully ensure robust target coverage in each fraction. The volume of SC that does not comply with the clinical goal adheres to the FLASH effect conditions in each fraction. Depending on the aforementioned fractionation schemes used, a FMF of approximately 0.6 to 0.8 is necessary to enable such treatment in FLASH conditions. \\   
\textit{Significance.} Our study demonstrates the potential of hypofractionated FLASH PT in treating spine metastasis while preserving SC integrity. This approach could enable more effective treatment of spinal metastases, potentially preventing SC compression and paralysis in these patients.
\end{abstract}

\keywords{Conformal FLASH \and Spine metastasis \and Proton therapy \and SBRT}

\section{Introduction}


Spinal metastases affect up to 70\% of terminal cancer patients \cite{hong_updated_2022} and can significantly impact their quality of life (QoL), primarily due to severe pain, decreased mobility, and potential neurological deficits \cite{liu_quality_2022}. The management of spinal metastases typically involves a multidisciplinary approach including surgery, radiation therapy (RT), and systemic treatments. RT is predominantly delivered using conventional external beam radiation therapy (cEBRT), aiming at pain palliation, prevention of local disease progression, and maintenance or recovery of neurological function \cite{mizumoto_radiotherapy_2011}. However, the maximal dose which can be delivered using cEBRT is often limited by the adjacent critical organs, such as the spinal cord, resulting in suboptimal local tumor control.

Specialized radiotherapy techniques, such as stereotactic body radiotherapy (SBRT), bring higher precision by delivering a high dose to the tumor while sparing surrounding healthy tissues \cite{glicksman_stereotactic_2020}. These advancements are particularly promising for patients with longer life expectancies and oligometastatic diseases, leading to potential improvements in local control and pain relief. However, in the case of metastases encompassing the spinal cord, SBRT is contraindicated given that spinal cord tolerance is no longer respected. 

The discovery of the FLASH effect represents another significant advancement in RT. FLASH is characterized by the sparing of organs at risk (OAR) through ultra-high dose rate (UHDR) irradiation without compromising tumor control \cite{vozenin_biological_2019}. In vivo studies have confirmed the advantage of FLASH radiotherapy in mouse \cite{favaudon_ultrahigh_2014,montay-gruel_irradiation_2017}, mini-pig and cat cancer models \cite{vozenin_advantage_2019}. When combined with proton pencil beam scanning (PBS), the feasibility of conformal FLASH PBS planning has already been demonstrated in several treatment sites \cite{wei_use_2022,kaulfers_pencil_2024,lattery_pencil_2023} through the use of a patient-specific range modulator \cite{kang_universal_2022,deffet_optimization_2023}. The high precision of protons, their sharp dose gradient, and the potential OAR-sparing effect of FLASH make this modality a potential candidate for treating indications previously untreatable with conventional dose rates (DR) using hypofractionated schemes.

In the context of spinal metastases, conformal FLASH PBS could therefore revolutionize treatment by enabling radical intervention for cases currently deemed non-treatable, thus preventing spinal cord compression, VCF, and paralysis. The study objective is therefore to demonstrate that FLASH-enabled Proton SBRT plans may be used as an alternative treatment for spinal metastases and establish the clinical conditions required to achieve this, in terms of both biological effectiveness and treatment delivery. 


\section{Materials and methods}
\subsection{Patient cohort and target delineation}

\begin{figure}[!h]
    \centering
    \includegraphics[width=0.5\linewidth]{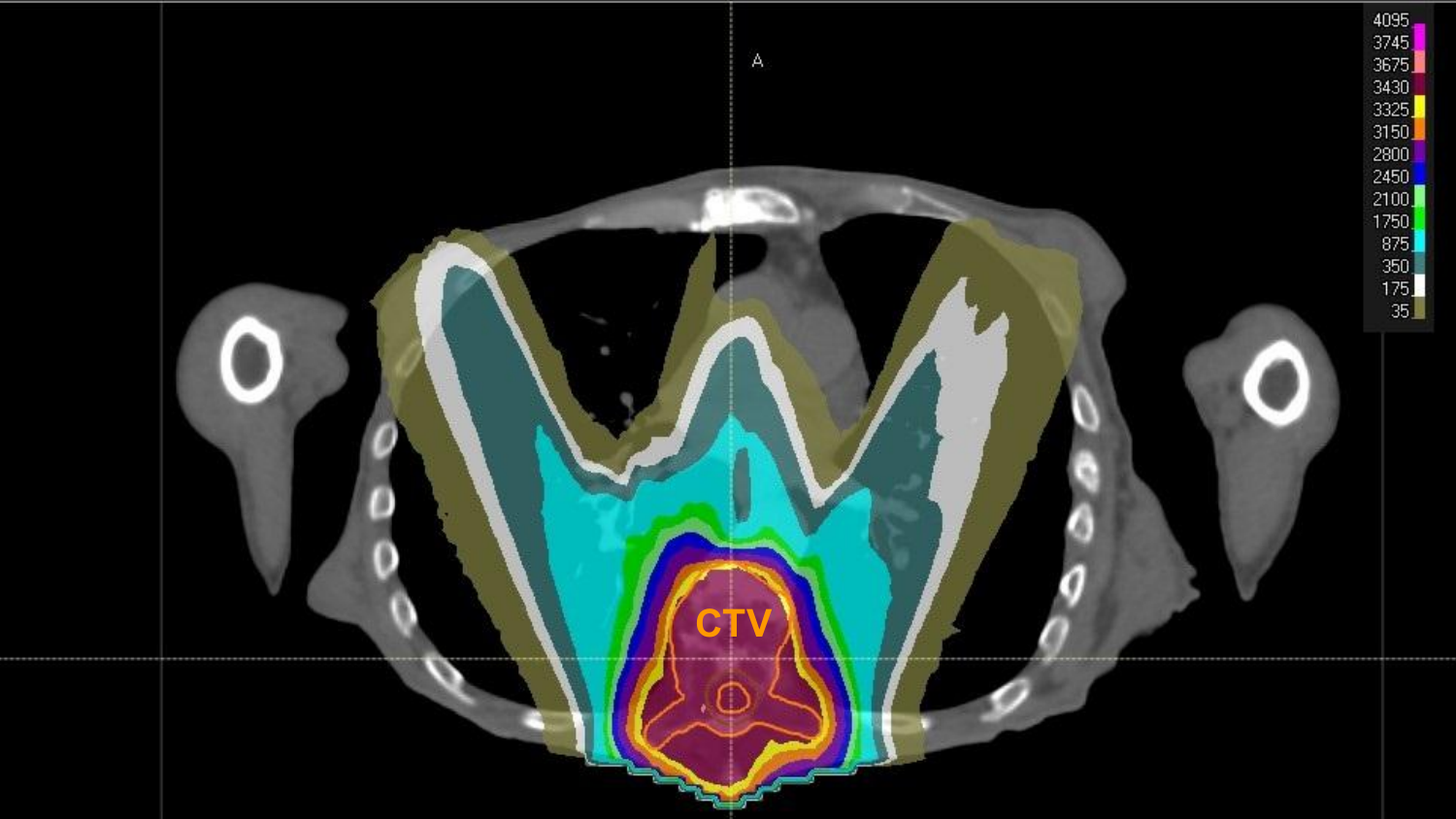}
    \caption{Isodoses for the 7 x 5 Gy fractionation scheme. The priority was on achieving the FLASH effect in the spinal cord. }
    \label{fig:vertebra}
\end{figure}

A single patient with spine metastasis was selected for this case study. The anonymized patient data was retrieved from a retrospective database from Cliniques Universitaires Saint-Luc
(Brussels, Belgium), for which ethical approval was granted. The clinical target volume (CTV) included the entire vertebra, encompassing the spinal cord. The spinal cord, as an organ at risk (OAR), was identified as the only critical structure that necessitated careful sparing (see Fig.~\ref{fig:vertebra}).
\subsection{Treatment planning strategy}
The conformal FLASH PBS proton plans were robustly optimized using a research version of RayStation 11B available exclusively for FLASH planning.

The beam configuration consisted of two lateral posterior obliques beams positioned at 160° and 200°. This configuration was chosen to ensure adequate target coverage. For each beam, a 6.5 slab aluminium range shifter was used, along with a 6 cm brass block. This guaranteed that the highest machine energy and thus maximum possible beam current could be reached. The brass block was conformed to provide a 1 cm margin around the CTV. The proton treatment machine used for the simulation is an IBA Proteus Plus. 

The robust optimization process included two stages: initial standard dose robust optimization followed by dose rate (DR) optimization. During robust optimization, single field optimization (SFO) was applied, meaning each beam was optimized separately according to the fractionation scheme (see section \ref{subsec:fractionation}). Moreover, the plan was designed with only one beam delivered per fraction, thus a certain number of fractions were assigned to each beam to deliver the total dose. Robust objective functions focused on ensuring minimum total coverage for the CTV only. The optimization included a total of 21 scenarios: 7 setup scenarios (1.5 mm shifts in the 6 different Cartesian directions, and the nominal case) times 3 range scenarios ($\pm 3\%$ range error, and the nominal case). For spot filtering, low-weight spots below a certain threshold were filtered (based on monitor units, MU). The threshold chosen for the minimum MU per spot in the plan, was case-specific (see next section). Spot sorting was optimized with a method, that finds the path among spots that maximizes the volume that receives FLASH, which in this case was the spinal cord.

\subsection{Fractionation regimens and clinical goals}
\label{subsec:fractionation}
To study the effect of different fractionation regimens on the FMF values, three regimes were tested: 42 Gy in 10 fractions (fx) of 4.2 Gy, 40 Gy in 8 fx of 5 Gy, and 35 Gy in 5 fx of 7 Gy.  The clinical goals for the target and spinal cord for each regimen are detailed in Table \ref{tab:Clinical_goals} in SM1 \cite{timmerman_story_2022}. 

SFO optimization was performed by assigning an even number of fractions per beam for the first two schemes. The third scheme was designed with 3 fractions for beam 1 and 2 fractions for beam 2. The minimum threshold for spot filtering was set to 22 MU per spot for the first two fractionation regimens and 34 MU per spot for the third one.

\subsection{Plan Evaluation}

The FLASH effect is assumed to be observed when the delivered dose exceeds 4 Gy and the dose rate surpasses 40 Gy/s \cite{ROTHWELL2022222}.

\subsubsection{Dose evaluation}

Target coverage and spinal cord clinical goals were evaluated using dose-volume histogram (DVH) metrics as standard. The nominal and robust CTV coverages were assessed both in total and per beam to ensure clinical goal compliance in each fraction.

\subsubsection{Dose rate evaluation}

Dose rate distribution per fraction and per beam was assessed for the spinal cord voxels receiving at least 4 Gy, by means of the dose rate-volume histogram (DRVH). The DVRH is based on the same concept as DVH, but using dose rate instead of dose. For evaluation, the 5\% percentile dose rate definition \cite{deffet_definition_2023} was used. 

\subsubsection{The FLASH-modifying factor (FMF)}

We used the FMF to quantify the difference in biological effectiveness between the dose delivered at UHDR ($D_\mathrm{UHDR}$) and at conventional dose rates ($D_\mathrm{CONV})$. It is defined by the ratio of the dose required to achieve the same biological effect under conventional conditions to the dose required under FLASH conditions  \cite{bohlen_normal_2022}, that is, 
\begin{equation}
    \mathrm{FMF} = \left. \frac{D_\mathrm{CONV}}{D_\mathrm{UHDR}}\right|_\mathrm{isoeffect} \enspace .
    \label{eq:FMF}
\end{equation}
A $\mathrm{FMF} < 1$ indicates greater normal tissue sparing for an equivalent level of tumor control. In practice, this means that the dose tolerance for OARs can be increased using the FLASH effect, allowing for a higher dose to be delivered to the tumor. This results in a higher probability of tumor control while preventing complications in normal tissues. The lower the FMF, the larger this effect becomes.

In this study, as FLASH plans are simulated, the dose distributions are converted into FMF-weighted doses with
\begin{equation}
    D_\mathrm{UHDR} = \mathrm{FMF}  \cdot D_\mathrm{CONV} \enspace .
\end{equation}

\section{Results}
\label{sec:results}
DVHs for the CTV for all regimens can be found in SM2 for each beam separately and a single fraction. Robust DVH bands are also computed. The clinical goal (D95$>$95\%Dp) is achieved for both beams in the nominal and the worst case scenario (see Table \ref{tab:results_CTV_fx}). This clinical goal is also achieved for the total dose in every case (34.5 Gy, 41.65 Gy and 39.63 Gy, respectively).

\begin{table}[h!]
\centering
\begin{tabular}{llllll}
\toprule
\multirow{2}{*}{\begin{tabular}[c]{@{}l@{}}Fractionation\\ regimen\end{tabular}} &
  \multirow{2}{*}{\begin{tabular}[c]{@{}l@{}}Goal per\\ fx {[}Gy{]}\end{tabular}} &
  \multicolumn{2}{c}{CTV D95\% Beam 1 {[}Gy{]}} &
  \multicolumn{2}{c}{CTV D95\% Beam 2 {[}Gy{]}} \\ \cline{3-6} 
               &      & \multicolumn{1}{c}{Nominal} & Worst Case & \multicolumn{1}{c}{Nominal} & Worst Case \\ \midrule
  
7 Gy x 5 fx    & 6.65 & \multicolumn{1}{c}{6.82}    & 6.77       & \multicolumn{1}{c}{6.81}    & 6.75       \\ 
5 Gy x 8 fx    & 4.75 & \multicolumn{1}{c}{4.84}    & 4.82       & \multicolumn{1}{c}{4.94}    & 4.89       \\ 
4.2 Gy x 10 fx & 3.99 & \multicolumn{1}{c}{4.1}     & 4.03       & \multicolumn{1}{c}{4.02}    & 3.97       \\  \bottomrule  
\end{tabular}%
\caption{CTV coverage results per beam, for all fractionation schemes, in both nominal and worst case scenarios}
\label{tab:results_CTV_fx}
\end{table}

\begin{figure}[h]
    \centering
    \includegraphics[width=0.7\linewidth]{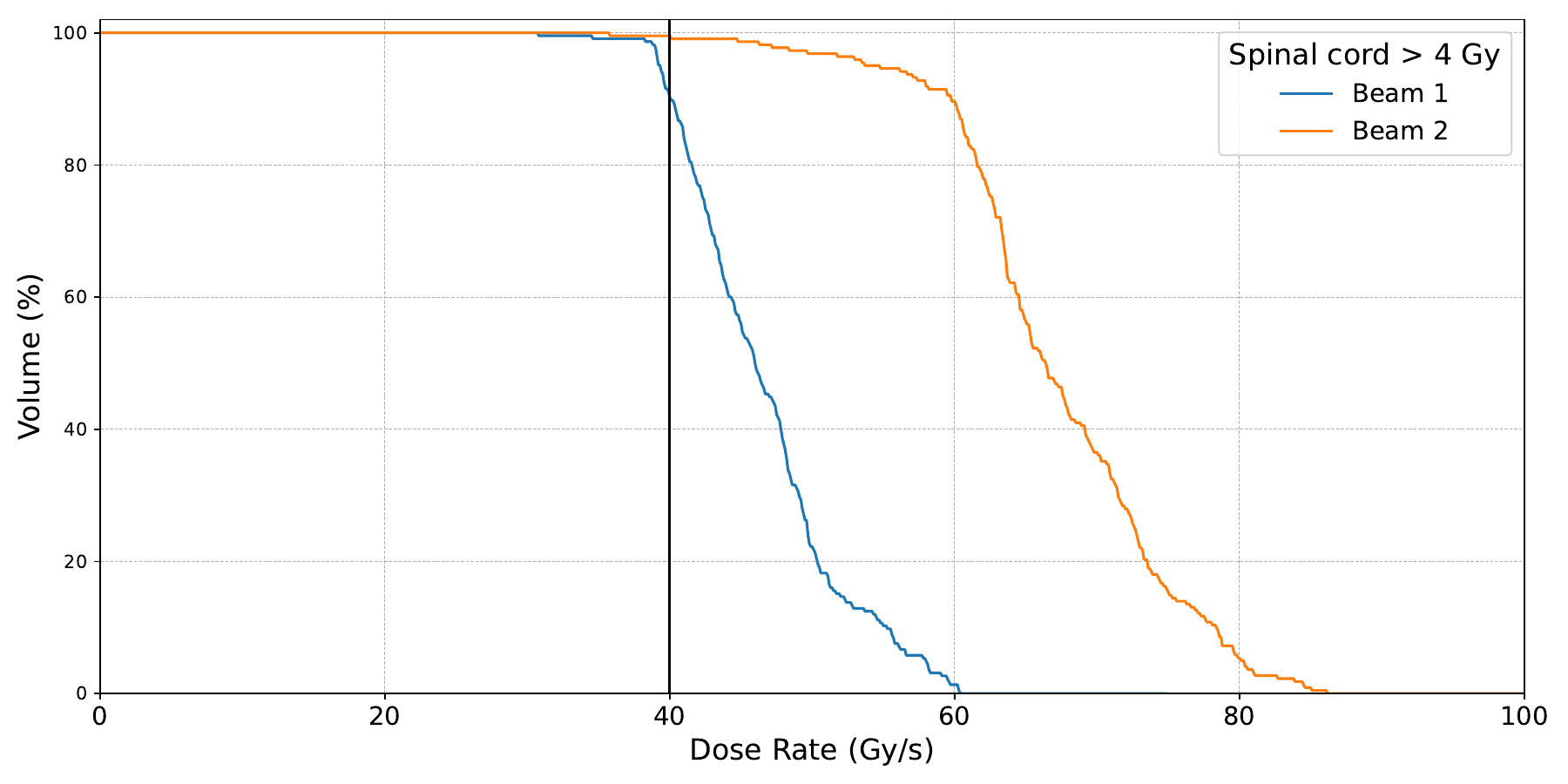}
    \caption{DRVH of spinal cord for each beam and per fraction (7 Gy x 5 fx regimen). UHDR (i.e. $>$ 40 Gy/s) is achieved for both beams in the high dose spinal cord region (i.e., $>$ 4 Gy).}
    \label{fig:drvh}
\end{figure}

Next, the DRVH for the spinal cord is presented in Fig. \ref{fig:drvh} for each beam separately within a single fraction. The plotted volume of the spinal cord corresponds to the region irradiated above 4 Gy, which is more likely for benefiting from a FLASH effect. The DRVH indicates that both beams achieve an UHDR in this spinal cord region, covering 98.60\% and 99.33\% of the volume with DR $>$ 40 Gy/s for beam 1 and beam 2, respectively.

\begin{figure}[h]
    \centering
    \includegraphics[width=\textwidth]{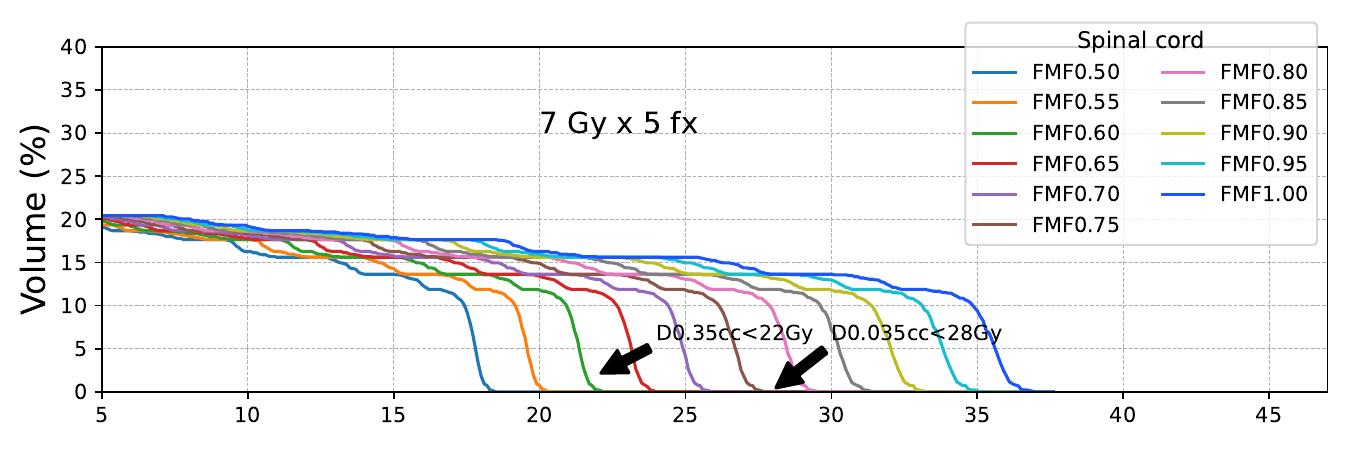}
    \caption{DVH of spinal cord re-weighted by flash modifying factor (FMF) values for the 5 fx of 7 Gy scheme.}
    \label{fig:fmf-main}
\end{figure}

To determine the necessary FMF value in the spinal cord to enable such treatment (i.e., meet the clinical goals) in FLASH conditions, the spinal cord DVH was re-weighted using a range of realistic values. Figure \ref{fig:fmf-main} illustrates the result alongside the clinical goals associated with the 5 fx of 7 Gy fractionation regimen. The other two regimens can be found in SM4 (Fig. \ref{fig:fmf-main}). The analysis shows that the FMF increases as the number of fractions increases (and the dose per fraction decreases), with values ranging from 0.6 to 0.8.

\section{Discussion}
In this study, proton SBRT plans were designed for a challenging spine metastasis case and subsequently converted to conformal FLASH plans within the RayStation TPS. The aim was to demonstrate that, based on the considered assumptions, the FLASH-enabled proton SBRT plan would allow for the radical treatment of lesions, which would not be treatable with conventional RT. This is particularly critical in cases such as the one presented in this study, where the metastasis encompasses the whole spinal cord, which has limited radiation tolerance and high toxicity risk. The spinal cord could therefore be spared by means of the FLASH effect, while ensuring proper tumor control.

The analysis shows that UHDR can indeed be achieved within this sensitive OAR, creating favorable conditions for the FLASH effect to occur. It also shows that the spinal cord FMF necessary to enable such treatment under FLASH conditions is highly dependent on the fractionation scheme. This dependence is expected, as decreasing the dose per fraction typically increases the tolerance of healthy tissues to radiation, a relationship that is well documented in a recent review study \cite{bohlen_normal_2022}. The referenced study also provides average FMF values and their standard deviations, reporting 0.95 $\pm$ 0.11 for all data below a single fraction dose of 10 Gy, with values decreasing to as low as 0.7 for mammalian skin-reaction data, but only above 25 Gy. Consequently, for spinal cord irradiated with less than 10 Gy per fraction, as in our study, the FMF of 0.95 suggests that none of the fractionation regimens we presented could comply with the clinical goals. Only when the FMF gets smaller than 0.8, using the scheme with the lowest dose per fraction, is the spinal cord sufficiently spared.  However, achieving UHDR while exceeding the hypothetical dose threshold to enable FLASH conditions becomes even more challenging with lower doses per fraction. Nevertheless, more pre-clinical data is still needed to determine precise FMF values that may relate to specific toxicity endpoints for a certain OAR. These values may fall within a clinically achievable range for dose delivery, offering better treatment options for patients. Once experimental studies converge on FMF values for different organs in vivo, they will guide the selection of appropriate fractionation schemes for delivering FLASH plans and help predict the sparing effects from delivering FLASH treatment.

Plan robustness is key to achieve a good quality dose delivery, handle uncertainties, and ensure proper target coverage. However, achieving fully robust treatments is challenging when planning with the FLASH modality. The requirement for multi-field plans to be delivered on a one-beam-per-fraction scheme necessitates not only overall plan robustness but also individual beam robustness for each fraction. Additionally, the less conformal dose distributions obtained with planning in FLASH mode can result in target coverage loss due to setup and range errors, which in turn reduces both V4Gy and V40Gy/s in the surrounding OARs. This reduction may potentially compromise the FLASH effect.   

The observation of the FLASH effect in the spinal cord in our study was predicated for a dose delivered above 4 Gy and the DR surpassing 40 Gy/s in each fraction, which are values the most commonly found in the literature \cite{vozenin_biological_2019,montay-gruel_irradiation_2017,kacem_understanding_2022}. The DRVH analysis for the spinal cord confirmed that these thresholds were met, with over 98\% of the spinal cord volume receiving a DR greater than 40 Gy/s. However, it is important to note that the thresholds for the FLASH effect are not universally agreed upon as it may actually vary depending
on irradiation conditions and biology (e.g., tissues and oxygenation states). Pre-clinical studies have shown varying thresholds for the onset and saturation of the FLASH effect, with reported values for the DR ranging from less than 10 Gy/s to over 280 Gy/s \cite{montay-gruel_irradiation_2017,kacem_understanding_2022,cunningham_flash_2021,ruan_irradiation_2021}. In practice, it means that achieving such high dose rates and doses within the spinal cord may not be feasible. Consequently, the FLASH sparing effect may have no realistic chance of being observed for this specific indication.

Additionally, all the presented results are based on the 5\% PBS dose rate. Although, there is currently no consensus on which dose rate definition should be used. Various definitions exist \cite{deffet_optimization_2023}, with some averaging the dose rate and others taking into account the time structure of spot delivery. It remains unclear which definition best correlates with the FLASH effect. This uncertainty complicates the design and interpretation of FLASH treatment plans and highlights the need for further research to establish a standardized and biologically relevant dose rate metric.




Finally, proton therapy, particularly with advanced delivery techniques like conformal FLASH PBS, involves significant financial costs \cite{yan_global_2023}. However, the potential benefits may be considerable for patients with spinal metastases which are not treatable with conventional RT. In this setting, the high precision and favorable dosimetry of protons, combined with the potential tissue-sparing effects of FLASH, may offer enhanced pain relief, better neurological function preservation, and reduced treatment-related morbidity. Therefore, the cost of such an advanced treatment modality must be well-balanced with the clinical endpoint for patients.


\section*{Acknowledgements}
Sophie Wuyckens and Macarena Chocan Vera benefited from a financial support of Wallonia in the frame of a MecaTech’s and BioWin’s Clusters program. John A.~Lee is a Research Director with the Belgian F.R.S.-FNRS.

\bibliography{main}
\bibliographystyle{unsrt}

\clearpage
\section*{Supplementary Material}
    
\beginsupplement

\section*{SM1. Clinical goals}
\begin{table}[h!]
    \centering
    \begin{tabular}{llll} 
    \toprule
         Clinical Goal&  4.2 Gy x 10 fx&  5 Gy x 8 fx&7 Gy x 5 fx\\
         \midrule
         CTV D95&  39.90 Gy&  38 Gy&33.25 Gy\\
         Spinal Cord D0.35cc&  -&  -&22 Gy\\
         Spinal Cord D0.03cc&  36 Gy&  33.6 Gy&28 Gy\\
 Spinal Cord V31Gy& 5 cc& -&-\\
 Spinal Cord V26.4Gy& -& 0.35 cc&-\\ \bottomrule
    \end{tabular}
    \caption{Clinical goals for CTV and spinal cord, according to the fractionation regimens studied (Timmerman, The Red Journal, 2022).}
    \label{tab:Clinical_goals}
\end{table}

\section*{SM2. Robustness results}

\begin{figure}
  \centering
  \subfloat[a][7 Gy x 5 fx]{\includegraphics[width=.7\linewidth] {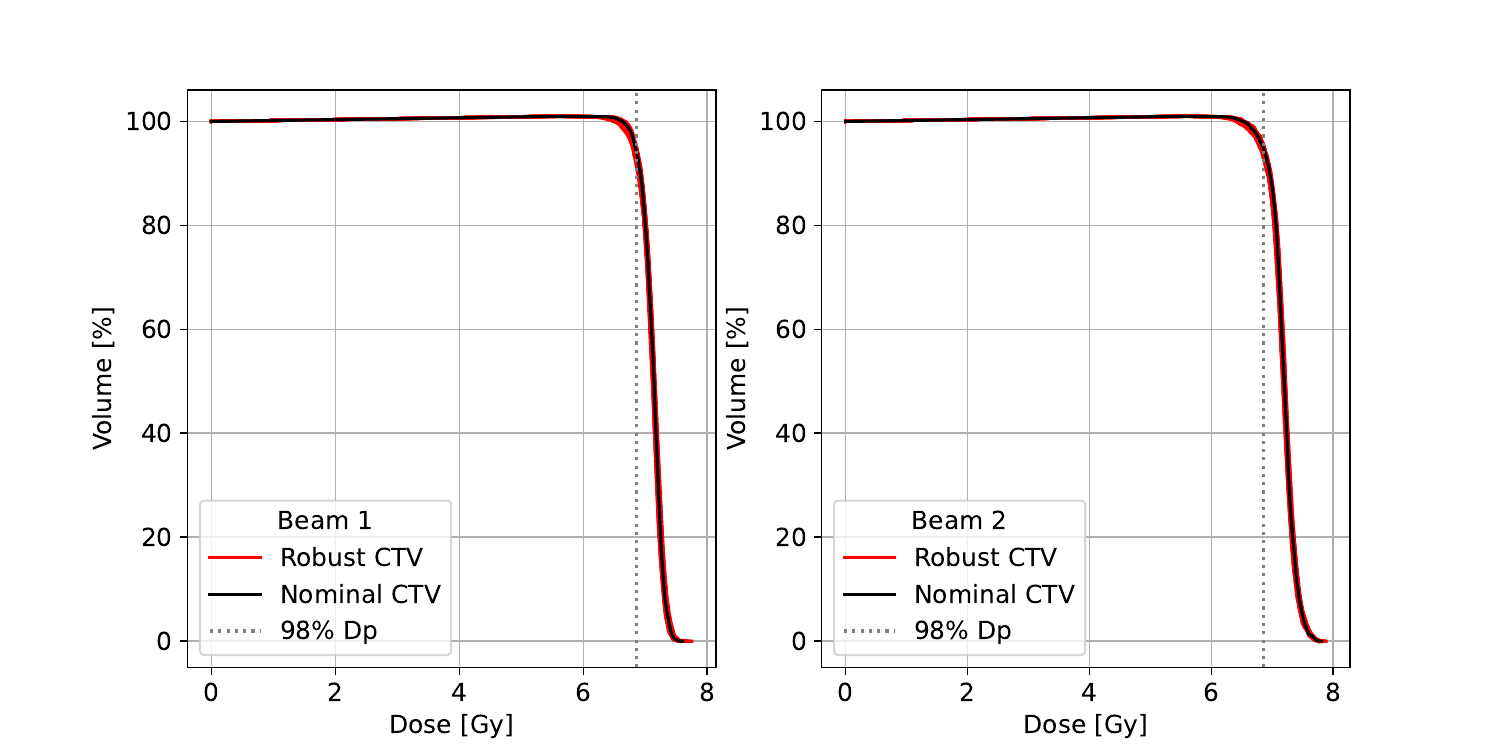} \label{fig:DVH7 Gy x 5 fx}} \\
  \subfloat[b][5 Gy x 8 fx]{\includegraphics[width=.7\linewidth] {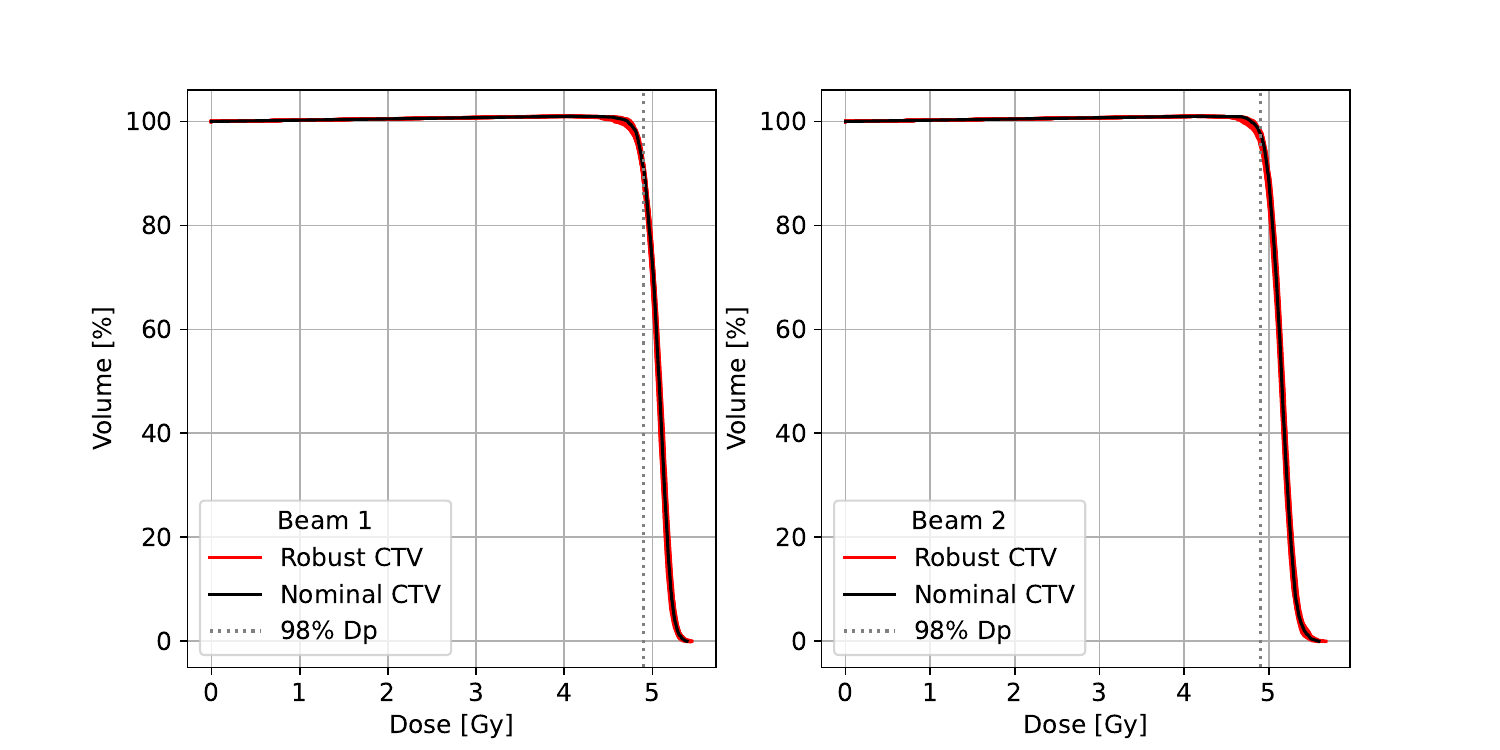} \label{fig:DVH5 Gy x 8 fx}}
  \\ 
  \subfloat[c][4.2 Gy x 10 fx]{\includegraphics[width=.7\linewidth] {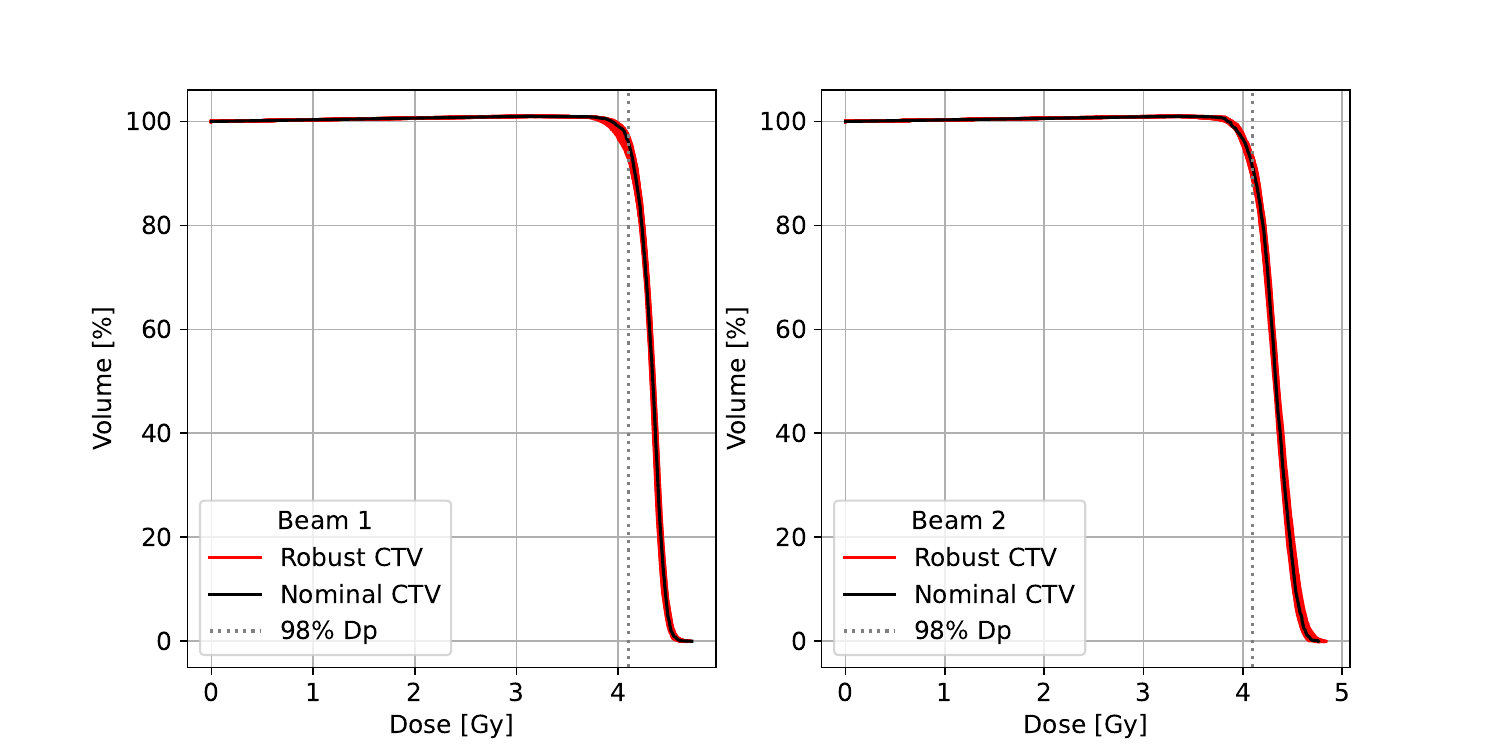} \label{fig:DVH4.2 Gy x 10 fx}}
  \caption{DVH of CTV for each beam and per fraction (all regimes). The nominal and robust CTV coverages are achieved for both beams in every case.} \label{fig:allfractionationsDVH}
\end{figure}

\newpage

\section*{SM3. Spinal Cord DRVH for non-clinical fractionation schemes}

\begin{figure}[h!]
  \centering
  \subfloat[a][5 Gy x 8 fx]{\includegraphics[width=.6\linewidth] {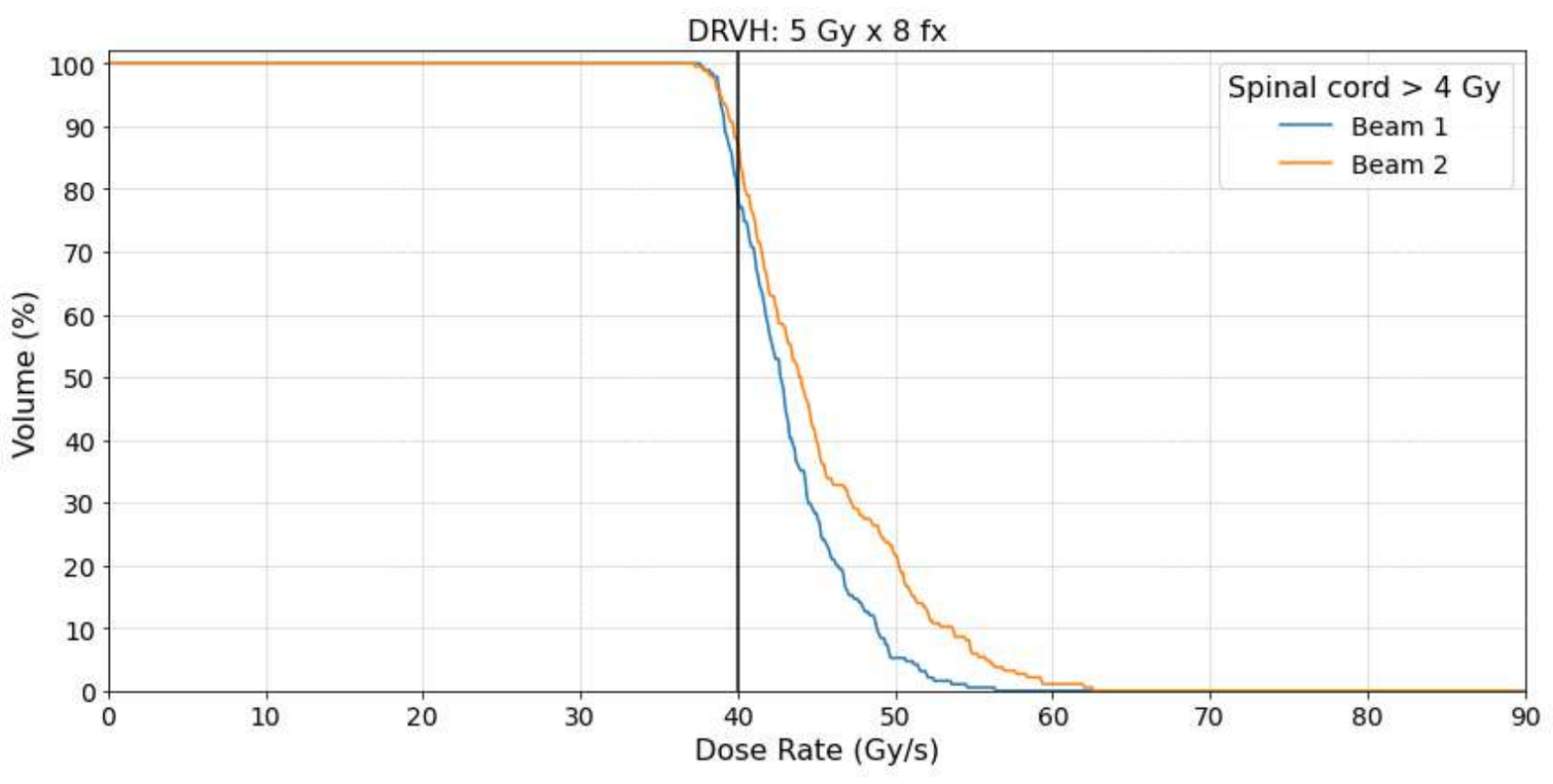} \label{fig: 5 Gy x 8 fx}} \\
  \subfloat[b][4.2 Gy x 10 fx]{\includegraphics[width=.6\linewidth] {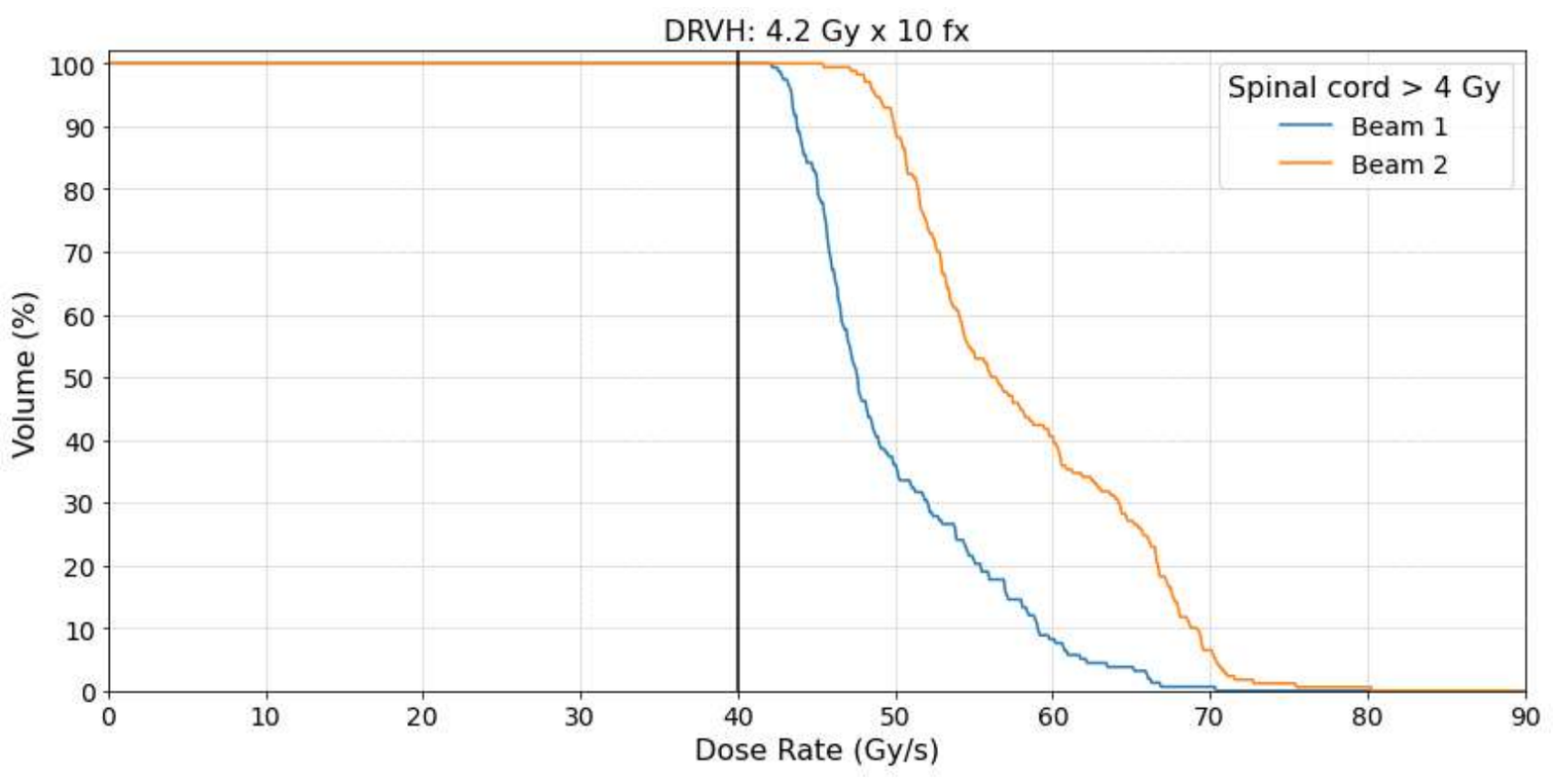} \label{fig: 4.2 Gy x 10 fx}}
  \caption{DRVH of Spinal Cord for each beam and per fraction, for the portion of volume receiving more than 4 Gy, for 2 different fractionation schemes. While (a) only achieves 40 Gy/s in 80\% of the volume, regime (b) achieves V40Gy/s = 100\% .} \label{fig:allfractionations}
\end{figure}

\newpage
\section*{SM4. Spinal Cord FMF-weighted DVH for non-clinical fractionation schemes}

\begin{figure}[h]
    \centering
    \includegraphics[width=\textwidth]{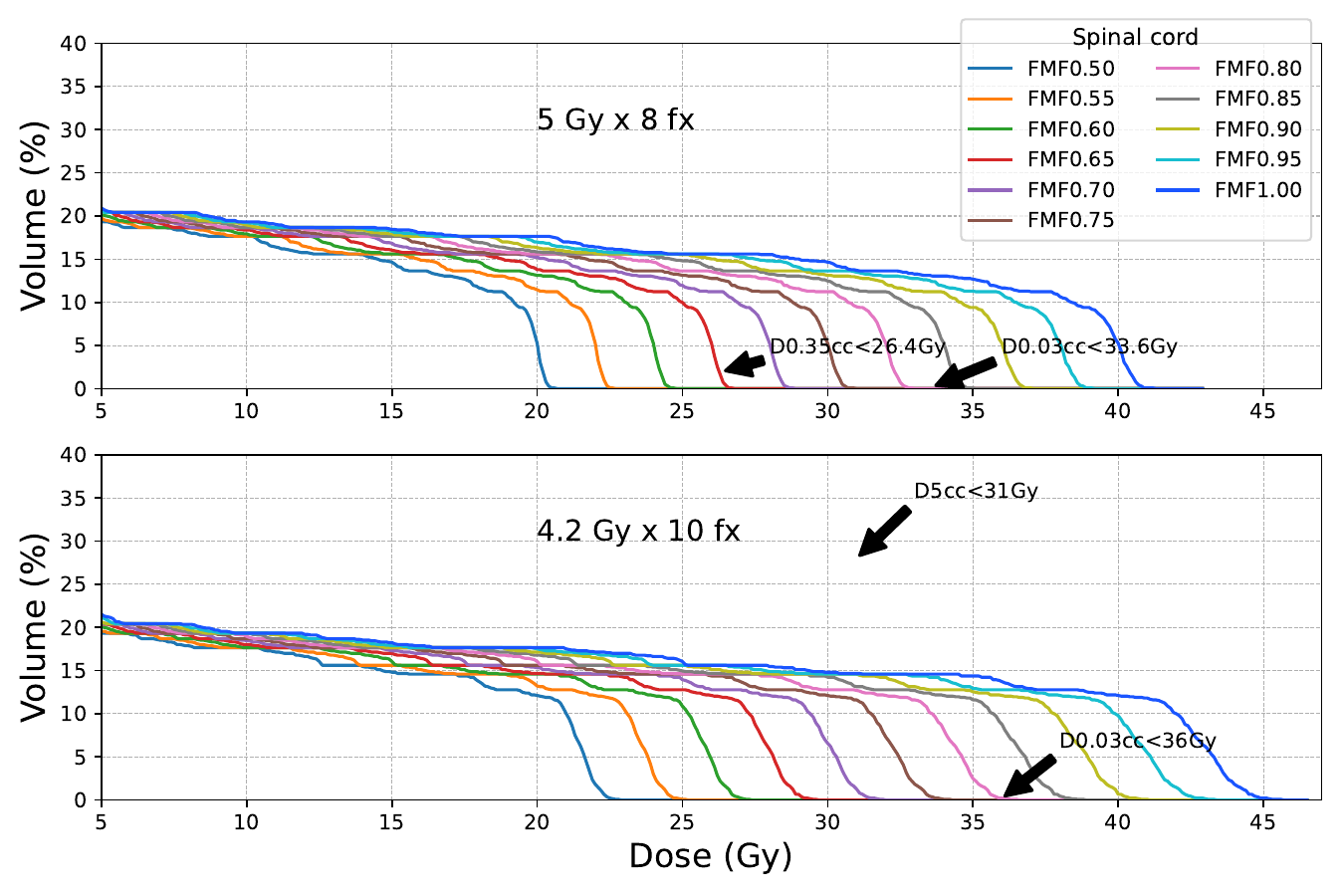}
    \caption{DVH of spinal cord re-weighted by flash modifying factor (FMF) values for 2 non-clinicql different fractionation schemes. The FMF required to comply with clinical goals increases as the number of fractions increases. }
    \label{fig:fmf}
\end{figure}

\end{document}